\documentclass[aps, superscriptaddress, secnumarabic, twocolumn, amsmath,amssymb,nofootinbib,nobalancelastpage]{revtex4}
\usepackage[mathscr]{eucal} 
\usepackage{graphicx}
\usepackage{longtable}
\usepackage{epstopdf} 

\newcommand{\vect}[1]{\boldsymbol {#1}}

\newcommand{\ti}[1]{#1}  
\newcommand{\vol}[1]{\bf #1}  

\allowdisplaybreaks[4]
\begin{document}

\title{Discussion of \\ Di Vita, A., ``On information thermodynamics and scale invariance in fluid dynamics'', {\it Journal of Thermodynamics \& Catalysis} 3: e108 (2012)}
\author{Robert K. Niven}
\email{r.niven@adfa.edu.au}
\affiliation{School of Engineering and Information Technology,
The University of New South Wales at ADFA, 
Canberra, ACT, 2600 Australia.} 

\author{Bernd R. Noack}
\email{bernd.noack@univ-poitiers.fr}
\affiliation{Institut PPRIME, 
CNRS -- Universit\'e de Poitiers -- ENSMA, UPR 3346, 
CEAT, 
F-86036 POITIERS Cedex, France.} 


\date{6 August 2013; 15 April 2014}




\maketitle
%
%

%

In a recent paper to the open-access journal {\it Journal of Thermodynamics \& Catalysis}, Di Vita \cite{DiVita_2012} makes two separate sets of comments: 
\begin{enumerate}
\item A critique of our joint work on a maximum-entropy (MaxEnt) closure of a Galerkin principal component decomposition of a periodic flow system \cite{Noack_Niven_2012}; 
\item Comments on the theoretical basis of the maximum entropy production (MaxEP) hypothesis for prediction of the steady state of a flow system, advocated by a number of authors \cite{Paltridge_1975, Paltridge_1978, Ozawa_etal_2003, Kleidon_L_book_2005, Martyushev_S_2006}. The latter is connected to a direct MaxEnt analysis of flow systems given by us, based on an entropy defined on the set of instantaneous flux states \cite{Niven_2009_PRE, Niven_2010_PTB, Niven_2012_AIP, Niven_Noack_2013_Springer}.
\end{enumerate}
Di Vita's first critique is incorrect, as discussed in detail below.  Furthermore, the short qualitative discussion given by Di Vita provides a misleading picture of the very purpose of applying entropic inference to the dynamical system examined.  
In particular, Di Vita suggests two ad-hoc scaling transformations of the closure 
which the underlying Galerkin system does not have, and which violate the physics of a self-amplified amplitude-limited oscillation. Scaling transformations should only be applied to closures 
if these transformations are consistent with the evolution equation 
which the closures aim to characterise. 

In the following, we respond to Di Vita's comments on the MaxEnt closure of the Galerkin system 
and on theoretical principles, both in \cite{DiVita_2012} and in \cite{DiVita_2010}.

\subsection*{Response to Di Vita's \cite{DiVita_2012} 
critique of the scaling argument used in \cite{Noack_Niven_2012}}

In \cite{DiVita_2012}, Di Vita makes a scale invariance argument to criticise the analysis given in \cite{Noack_Niven_2012}.  However, we are afraid that he has misinterpreted our analysis. As defined in eq.\ (2.1) of \cite{Noack_Niven_2012}, our $a_i$'s are dimensionless, representing dimensionless coordinates in a velocity phase space.  Only the expansion modes $\vect{u}_i$ have dimensions of velocity before the non-dimensionalisation. Di Vita's argument, based on dimensional $a_i$'s, is therefore incorrect.  It is possible that Di Vita has misinterpreted our analysis in section 5, where we consider the ``pseudo" dimension of each quantity, to interpret the physical meaning of the Lagrangian multipliers. We had believed that this representation was clear (and it is separate to the rest of the paper).  

\subsection*{Response to Di Vita's \cite{DiVita_2012} choice of scaling for the maximum-entropy closure of a Galerkin system \cite{Noack_Niven_2012}}

In addition, Di Vita \cite{DiVita_2012}
questions the soundness of the maximum-entropy closure  of the Galerkin system \cite{Noack_Niven_2012}
by imposing a curious scaling argument on the first two ordinary differential equations
of the Galerkin system discussed in \cite{Noack_Niven_2012} (see also \cite{Noack_Niven_2013}):
\begin{subequations}
\label{Eqn:GS}
\begin{eqnarray}
\dot a_1 &=& \sigma_1 a_1 - \omega_1 a_2 + h_1, \\
\dot a_2 &=& \sigma_2 a_2 + \omega_1 a_1 + h_2,
\end{eqnarray}
\end{subequations}
where $h_1$ and $h_2$ are higher-order terms.
These two equations evidently describe an oscillator.
For the post-transient periodic solution, 
which this study aims to approximate,
the growth-rate and nonlinear terms effectively cancel each other 
in a one-period average.
This behaviour is described and employed in numerous articles by the first author
\cite{Noack_EtAl_2003, Noack_Morzynski_Tadmor_2011,Tadmor_EtAl_2011}
and detailed in textbooks of oscillatory dynamics.
The correspondingly filtered ordinary differential equations read:
\begin{subequations}
\label{Eqn:HarmonicOscillator}
\begin{eqnarray}
\dot a_1 &=& - \omega_1 a_2, \\
\dot a_2 &=& + \omega_1 a_1.
\end{eqnarray}
\end{subequations}
Modulo a phase shift, 
the solution describes the phase-invariant limit cycle
with frequency $\omega$ and radius $R$:
\begin{subequations}
\label{Eqn:LimitCycle}
\begin{eqnarray}
\label{Eqn:LimitCycle:1}
a_1 &=& R \> \cos \omega_1 t, \\
a_2 &=& R \> \sin \omega_1 t.
\label{Eqn:LimitCycle:2}
\end{eqnarray}
\end{subequations} 
This harmonic motion approximates well \eqref{Eqn:GS}
when the radius $R$ is determined from energetic (or other) considerations.
There exists only one non-trivial radius $R$ for which the average power vanishes.
Curiously, Di Vita imposes a parabola (!) as the scaling relationship
\begin{equation}
\label{Eqn:ScalingRelationship}
a_2 = \alpha a_1^2.
\end{equation}
The source of this relationship is called `dimensional argument', but pretty much left in the dark.
He argues further that the MaxEnt closure is not invariant with respect to this imposition,
hence the MaxEnt closure must be flawed.

We argue that a parabola is an arbitrarily poor approximation of a circular limit-cycle,
is inconsistent with the effective phase-invariance property of the Galerkin system,
and should hence not even be considered as a scaling relationship.
If \eqref{Eqn:ScalingRelationship} were true, 
then $a_2$ would either be always non-negative or always non-positive, violating \eqref{Eqn:LimitCycle:1}.
Starting with \eqref{Eqn:LimitCycle:2}, 
we obtain $a_2 = \alpha \left( 1 + \cos 2 \omega_1 t \right) /2$ 
which is obviously incompatible with \eqref{Eqn:LimitCycle:2}.
Even if the approximate solution of the Galerkin system was unknown,
the equations are effectively phase-invariant and mirror symmetric for $a_1,a_2$.
If $a_2 = \alpha a_1^2$ is true, 
then $a_2= -\alpha a_1^2$, $a_1 = \alpha a_2^2$, etc, would be true as well,
imposing $a_1=a_2=0$ as the only solution.
However, $a_1=a_2=0$ is in contradiction 
to the assumed non-trivial limit cycle behaviour.
Even if the Galerkin system was not effectively phase invariant,
why should a transformation impose a single sign on one of the amplitudes?
Principal orthogonal decomposition (POD) modes should vanish on average,  leading again to the unphysical solution $a_1=a_2=0$.
Hence, the curious scaling relationship \eqref{Eqn:ScalingRelationship} should not be applied.
Thereafter, the construed counter-argument collapses.

In addition, Di Vita also requests scaling invariance $E \mapsto \lambda E$ from the closure,
this time without going into the analytics.
Why should the MaxEnt closure be indifferent to arbitrary energy levels
when the Galerkin solution and the resulting power equation 
permit only one non-trivial fluctuation level corresponding to von K\'arm\'an vortex shedding?
Again, a symmetry is suggested which is in contradiction to the nature
of nonlinear oscillation with a stable limit cycle.
Scaling arguments should only be applied to closures 
when the constitutive evolution equations have the scaling properties.
Clearly, Di Vita's criticism is based on a misunderstanding 
of the foundation of such scaling arguments.

\subsection*{Broader comments on theory \cite{DiVita_2012, DiVita_2010}} 

We also wish to raise some broader questions concerning nonequilibrium thermodynamics and the analysis of dissipative systems.  Firstly, we are curious why Di Vita pays so much attention to Prigogine's \cite{Prigogine_1967} MinEP principle?  For a system with a unique steady state -- which generally implies a system which is ``near equilibrium''  -- this merely identifies the steady-state solution in the space of transient non-solutions. The steady-state solution is usually also obtainable by direct calculation, in which case Prigogine's principle tells us nothing new. Indeed, we are unaware of any example in any branch of engineering in which Prigogine's principle  is actually invoked.  We here distinguish Prigogine's principle from other MinEP or related principles which have found utility in engineering, such as the MinEP engineering design principle of Bejan \cite{Bejan_1996}, the MinEP limit given in finite-time thermodynamics \cite{Salamon_A_G_B_1980, Salamon_B_1983, Nulton_etal_1985} or its steady-state analogue \cite{Niven_Andresen_2009}, or the minimum power principle invoked in the analysis of electrical networks \cite{Jeans_1966, Landauer_1975}.


Prigogine's principle is fundamentally different to the entropy production extremum hypotheses / principles of interest to us and many others, especially the MaxEP principle advocated by Paltridge \cite{Paltridge_1975} and its inversion to a MinEP principle by switching between flux and force constraints \cite{PaulusJr_2000, PaulusJr_G_2004, Martyushev_2007, Niven_2010_JNET, Kawazura_Y_2010, Kawazura_Y_2012}. Such principle(s), if they exist, would identify the observable steady state from the set of many possible steady states. This would provide (or extract) some extremely useful information, without the need for a full dynamical solution. This utility explains the strong attention this topic has received in the literature.

Di Vita \cite{DiVita_2012} dismisses the existence of the Paltridge MaxEP principle on the grounds that it ``involves the introduction of additional hypotheses, which by themselves are less evident than [MaxEP] itself.'' This criticism may be correct insofar as it applies to Ziegler's orthogonality principle \cite{Ziegler_1977} and to early theoretical treatments of the Paltridge MaxEP principle \cite{Dewar_2003, Dewar_2005}, in which errors have been identified \cite{Bruers_2007d, Grinstein_L_2007}. It could also be made of a recent treatment \cite{Dewar_M_2013} in which an irreversibility function -- itself the consequence of MaxEnt analysis -- is imposed as a constraint, which might be considered to invoke circular reasoning. 
Di Vita's criticism could also be levelled at the various formulations of upper bound theory in turbulent fluid mechanics \cite{Malkus_1956, Busse_1970, Kerswell_2002}, in which the postulated extrema may appear rather {\it ad hoc}.  However, other theoretical treatments have been proposed to explain entropy production extrema, including our above-mentioned MaxEnt analysis of an infinitesimal flow system \cite{Niven_2009_PRE, Niven_2010_PTB, Niven_2012_AIP, Niven_Noack_2013_Springer}.  The analysis is based on maximisation of an entropy defined on the set of instantaneous flux states and reaction rates, giving a potential function (negative Massieu function) which is minimised at steady-state flow.  In certain circumstances, this can be interpreted to give rise to a ``secondary'' MaxEP principle.  We do not believe that this analysis invokes any additional hypotheses; indeed, it preserves the Legendre mathematical structure also evident in equilibrium thermodynamics. It cannot be so readily dismissed by Di Vita.

Furthermore, Di Vita \cite{DiVita_2012} comments that he has criticised the first author's 2009 paper \cite{Niven_2009_PRE} in his 2010 paper \cite{DiVita_2010}. However, the only thing we can find is an oblique comment on our choice of prior probabilities, and that the analysis does not give a MaxEnt equivalent of  Liouville's equation. These comments are so vague that we do not see how to interpret them as a criticism. In the first study \cite{Niven_2009_PRE} we do adopt a uniform prior probability, but in later works \cite{Niven_2010_PTB, Niven_2012_AIP, Niven_Noack_2013_Springer} we consider a prior probability in its full generality, which must be selected for the problem at hand. This is a feature of all MaxEnt analyses \cite{Jaynes_2003}.  Indeed, in the cited MaxEnt closure of a Galerkin decomposition \cite{Noack_Niven_2012, Noack_Niven_2013}, solution requires the use of a non-uniform prior to account for marginal stability of the fixed point of the limit cycle.  The MaxEnt method is therefore tightly connected to Bayesian inference, in that it is necessary to choose (or infer) a prior probability from the problem specification. 

Concerning Liouville's equation, the idea of a mechanical basis of thermodynamics is certainly very old but quite flawed, especially if the $p$'s and $q$'s are insufficient to describe the system (e.g. for interacting particles, or in dissipative systems, or in flow systems under the Eulerian description). Callen \cite{Callen_1985}, for example, does not require or even mention Liouville. Even the masterful treatise by Hill \cite{Hill_1956} examines Liouville, but does not actually use it. Writing Boltzmann's principle (``MaxProb") \cite{Boltzmann_1877, Planck_1901} in the weak or inferential form \cite{Niven_2009}:
\begin{quote}
``{\it A system may be inferred to be in its most probable state}'',
\end{quote}
we see that this principle applies to {\it any} probabilistic system, regardless of whether it is a thermodynamic system or involves thermodynamic equilibrium. Indeed, this principle provides the foundation of Large Deviations theory in statistical inference \cite{Ellis_1985, Touchette_2009} and also of the Method of Types in communication theory \cite{Csiszar_1998}. To the extent that the relative entropy function and Jaynes' MaxEnt method are derivable from Boltzmann's principle, they also apply to any probabilistic system regardless of its connection to thermodynamics.  We invoke this MaxEnt method of inference directly; no further dynamical principle is needed.  

Other than this, we  appreciate Di Vita's 2010 paper \cite{DiVita_2010} and its classification scheme. We  have two major points of departure, however:

\begin{enumerate}
\item Di Vita assumes local thermodynamic equilibrium (LTE), a very strong assumption, which strongly constrains his results. In chemically reactive systems, especially in natural systems, we {\it know} that LTE is not correct. The analysis then becomes messy, requiring empirical kinetic rate laws with fitted parameters, but in many systems this cannot be avoided. We also invoke LTE in many of our studies (primarily to define local temperatures, chemical potentials, etc), but do not see it as critical to the analysis; we would gladly discard it if a better method were available.

\item Although Di Vita \cite{DiVita_2010} gives a nice structure and some useful inequalities, expressing a number of variational principles with which to determine system stability -- indeed, which play the role of  constraints -- is it enough? Are there any other principles -- derived directly from probabilistic inference -- which can be used for system closure? If such principle(s) do exist, they will one day be considered of equal status to the four known laws of thermodynamics. Indeed, they will come to define ``thermodynamics'' as distinct from present-day ``thermostatics''.
\end{enumerate}

\vspace{10pt}
We thank Di Vita for his interest in our work.

\vspace{5pt}
\subsection*{Comment on {\it Journal of Thermodynamics \& Catalysis}} 

We also must add some commentary on the {\it Journal of Thermodynamics \& Catalysis}: we naturally attempted to respond to Di Vita's criticisms through a discussion paper in that journal.  However, after submission we received a poorly typeset proof of the manuscript.  We have never seen such poor handling of a manuscript by any journal - indeed one of our names was spelt incorrectly!  We did not consider it worthwhile to correct such a proof ourselves, and sought to have it re-typeset - however, we believe this proof went through to publication without our copyright permission or consent.  Despite several entreaties from us, the  journal does not appear to have attempted to make any correction.  Examining their website (http://omicsonline.org), the {\it Journal of Thermodynamics \& Catalysis} does not appear to have an Editor or any scientific oversight, although at time of writing the website appears to be malfunctioning and does not load any hyperlinks.  

Our response to Di Vita's article is outlined correctly herein. We hereby disassociate ourselves from any version of this manuscript published by the {\it Journal of Thermodynamics \& Catalysis}.


%

\end{document}